\documentclass[
  reprint,
  superscriptaddress,
  amsmath,amssymb,
  aps,
  prc, 
  floatfix,
  nofootinbib
]{revtex4-2}
\usepackage[utf8]{inputenc}
\usepackage{graphicx}
\usepackage{mathrsfs}
\usepackage{makecell}
\usepackage[caption=false]{subfig}
\usepackage{color}
\usepackage[dvipsnames]{xcolor}
\usepackage{comment}
\usepackage{ulem}

\def\sec@upcase#1{#1}
\usepackage{textcase}

\makeatletter
\def\@hangfrom@section#1#2#3{#1#2#3}

\def\sec@upcase#1{#1}

\makeatother

\begin{document}

\title{ \textit{Ab initio} \boldmath{$n\alpha$} scattering using the Efros method}
\date{\today} 

\begin{abstract}
We present the first \textit{ab initio} many-body application of the Efros method for scattering utilizing the harmonic oscillator expansion of the continuum wave function.  We select single-channel $n\alpha$ scattering and, using low-lying eigenstates from no-core shell model (NCSM) calculations, we obtain well-converged scattering results that are in reasonable agreement with experiment.  The results demonstrate that the Efros method is computationally practical and promising for further \textit{ab initio} many-body studies of nuclear reactions.

\end{abstract}
\author{Mamoon A. Sharaf}
\affiliation{Institute of Modern Physics, Chinese Academy of Sciences, Lanzhou 730000, China}
\author{Weijie Du}
 \email{Corresponding author: duweigy@gmail.com}
 \affiliation{Institute of Modern Physics, Chinese Academy of Sciences, Lanzhou 730000, China}
 \author{Peng Yin}
  \email{Corresponding author: pengyin@iastate.edu}
\affiliation{College of Physics and Engineering, Henan University of Science and Technology, Luoyang 471023, China}
\affiliation{State Key Laboratory of Heavy Ion Science and Technology, Institute of Modern Physics, Chinese Academy of Sciences, Lanzhou 730000, China}
\affiliation{Institute of Modern Physics, Chinese Academy of Sciences, Lanzhou 730000, China}
\author{Andrey M. Shirokov}
\affiliation{Skobeltsyn Institute of Nuclear Physics, Lomonosov Moscow State University, Moscow 119991, Russia}
\author{James P. Vary}
\affiliation{Department of Physics and Astronomy, Iowa State University, Ames, Iowa 50011, USA}
\maketitle

\section{Introduction}

Various {\it ab initio} (from first principles) methods have been developed and extensively applied to describe the structure of light nuclei at low energies.  These applications are a modern trend in the theory of light nuclei that are well-developed and have been successfully used to calculate various bound systems with the help of modern supercomputing techniques \cite{LeidemannOrlandini2013}.  As a result, those approaches have become celebrated methods for nuclear structure and investigations.  However, the situation with extending {\it ab initio } methods to calculations of nuclear reactions for systems with $A > 4$ nucleons is less developed \cite{CWJohnson}.  This is a pressing issue because of the need to develop \textit{ab initio} theories for investigations of astrophysical reactions, exotic and unstable nuclei, and various areas of applied physics.

There are extensive efforts to develop \textit{ab initio} reaction theory for systems with five or more nucleons based on the no-core shell model (NCSM) \cite{NCSMPRL, NCSMProg}.  The NCSM provides a convenient framework for accurate nuclear structure calculations.  The \textit{ab initio} methods that solve nuclear reactions based on the NCSM include the no-core Gamow shell model (NCGSM) \cite{Papadimitriou2013}, the resonating group method in combination with the NCSM (RGM/NCSM) \cite{QuagOne} and its extension, the NCSM with continuum (NCSMC) \cite{BaroniOne}, and the single-state harmonic oscillator representation of scattering equations (SS-HORSE) \cite{AMShirokovSSHORSE2016, SSHORSEIAMazur2017}, which is a simplification of the HORSE method \cite{Zaitsev1998,BangHORSE}.  We have applied the HORSE method to the reaction $np \rightarrow d\gamma$ \cite{DuWeijieHORSE}.

Despite the successes of these approaches in describing some nuclear reactions, significant computational challenges remain for \textit{ab initio} scattering calculations in systems with $A>4$ \cite{CWJohnson}.  In particular, the NCGSM, the RGM/NCSM, and the NCSMC substantially increase the computational cost of the shell-model calculations, as they are computationally demanding and difficult to scale to larger many-body spaces.  On the other hand, the SS-HORSE method is simple and can be used to obtain some important observables in light nuclear systems.  However, it does not provide the scattering wave functions required for calculating observables such as electromagnetic transition cross sections.  In addition, extending the SS-HORSE method to multichannel problems requires additional development.

To address these limitations, the Hulthén-Kohn method was modified by Efros in Ref. \cite{VDEfros2019} with the goal of facilitating its extension to continuum calculations within the NCSM.  In Ref. \cite{MethodsPaper2025}, we further modified the Efros method by introducing expansions in series of harmonic oscillator functions throughout, including the expansion of scattering wave functions as is done in the HORSE method.  In Ref.~\cite{M1E1arXiV2026}, we have applied the method we introduced in Ref. \cite{MethodsPaper2025} to solve $np \leftrightarrow d\gamma$ reactions up to $E_{\gamma}=20$ MeV.  This modified approach of \mbox{Ref. \cite{MethodsPaper2025}}, hereafter referred to as ``Efros method'', simplifies its use in combination with the NCSM which utilizes the basis of many-body oscillator functions.

The Efros method provides good convergence with a limited set of short-range functions (SRFs) comprised of oscillator functions with limited quanta accessible to modern no-core shell-model codes like MFDn \cite{MFDnMaris2010, Euro-Par, Aktulga, NCSM-NTSE2014, Shao}.  Therefore, it provides a practical way to extend the NCSM to reaction calculations.  The Efros method bypasses the limitation of the HORSE method, which in principle, can be used to calculate all scattering observables but is impractical for \textit{ab initio} many-body applications because of its prohibitive computational cost.  In addition, by combining the NCSM with the continuum wave function expansion in series of oscillator functions, the Efros method avoids computational difficulties encountered in the NCGSM, the RGM/NCSM, and the NCSMC. Unlike the SS-HORSE method, the Efros method provides both the scattering phase shift $\delta_{l}(E_{\textrm{rel}})$ and the scattering wave function, treating the long- and short-range components of the scattering wave function on an equal footing.  The method can be extended to multichannel problems \cite{VDEfros2019, MethodsPaper2025}.

In this work, we present the first \textit{ab initio} many-body application of the Efros method.  In particular, we investigate $n\alpha$ scattering, which provides a useful benchmark for \textit{ab initio} many-body approaches to low-energy nuclear reactions (see, for example, Refs. \cite{JCarlson1987,Nollett2007,Navratil2010,Papadimitriou2013,Lynn2016,AMShirokovSSHORSE2016, SSHORSEIAMazur2017, Lazauskas2018, AMShirokovSSHORSE2018,Kravvaris2020,Flores2025,Yang2025,Elhatisari2025,HangYu2026}).  We demonstrate that the Efros method is efficient and describes well $n\alpha$ scattering using a realistic $NN$ interaction, namely Daejeon16 \cite{Daejeon16}, establishing its prospect for future \textit{ab initio} studies of nuclear reactions.  We consider single-channel scattering at relative energies \mbox{$E_{\textrm{rel}} \leq 17$ MeV} with respect to the $n \alpha$ threshold.  We study the scattering phase shifts $\delta_{l}(E_{\textrm{rel}})$, cross sections $\sigma(E_{\textrm{rel}})$, resonance energies $E_{r}$, and widths $\Gamma$ in partial waves with spin-parities $J^{\pi}=3/2^{-}, 1/2^{-}, 1/2^{+}$.  We note that extending the Efros method to the scattering of charged particles is straightforward; see Refs. \cite{Yanikov2025,YanikovEfros2025}.

\section{Formalism}

Within the Efros method, the wave function describing elastic scattering of a neutron by a nucleus with the mass number $A$ at total center-of-mass (c.m.) energy $E$ is expressed in a single channel as \cite{VDEfros2019,MethodsPaper2025}
\begin{equation}
\Psi^{\Upsilon}=\frac{i}{2}[\eta^{-}-\eta^{+}\mathbb{S}(E_{\textrm{rel}})]+\sum_{q=1 }^{v}b_{q}(E)\beta_{q}.
\label{eq1}
\end{equation}
Here, the multi-index $\Upsilon=\{J, I_{1}, I_{2}, S, l\}$ specifies a partial wave.  It includes quantum numbers $J$, the total angular momentum; $I_{1}=1/2$, the spin of the neutron; $I_{2}$, the spin of the target nucleus; $l$, the orbital angular momentum of the $n\text{-}A$ relative motion; and $S$, the total channel spin.  $\eta^{\pm}$ describe the long-range behavior of the scattering wave function, $\beta_{q}, \hspace{1mm} q=1,...,v$ are $v$ linearly-independent ($A+1$)-body SRFs vanishing as $n\text{-}A$ separation $\xi_{A}\rightarrow \infty$ which are introduced to describe the wave function at short distances, and the $S$ matrix $\mathbb{S}(E_{\textrm{rel}})$ that is dependent on the energy of $n\text{-}A$ relative motion, $E_{\textrm{rel}}$, together with $v$ energy-dependent coefficients $b_{q}(E), \hspace{1mm} q=1,...,v$, comprise a set of $v+1$ unknowns to be determined.  For a single channel, $\mathbb{S}(E_{\textrm{rel}})=e^{2i\delta_{l}(E_{\textrm{rel}})}$, where $\delta_{l}(E_{\textrm{rel}})$ is the scattering phase shift.   $E_{\textrm{rel}}=E-E_{A}$, where $E_{A}$ is the ground state energy of the target nucleus.

The channel functions $\eta^{\pm}$ are expressed \cite{VDEfros2019}:
\begin{equation}
\eta^{\pm}=\hat{\mathcal{A}}[[\chi^{\frac{1}{2}}\Psi^{I_{2}}(\boldsymbol{\xi})]_{S}Y_{l}(\boldsymbol{\hat{\xi}}_{A})]_{JM_{J}}\tilde{h}^{\pm}_{l}(k\xi_{A})\Psi_{000}(\boldsymbol{R}_{\textrm{c.m.}}).
\label{eq2}
\end{equation}
The brackets $[...]_{S}$ and $[...]_{JM_J}$ denote angular momenta couplings to the channel spin $S$ and to the total spin $J$ and its projection $M_{J}$, respectively.  $\hat{\mathcal{A}}$ is the inter-fragment antisymmetrizer, $\chi^{\frac{1}{2}}$ is the neutron spinor, $\Psi^{I_{2}}(\boldsymbol{\xi})$ is the ground state wave function of the target nucleus in Jacobi coordinates \mbox{$\boldsymbol{\xi}=\{\boldsymbol{\xi}_{1}, \boldsymbol{\xi}_{2},...,\boldsymbol{\xi}_{A-1}\}$, and $Y_{l}(\boldsymbol{\hat{\xi}}_{A})$} is the spherical harmonic.  $k=\sqrt{2 m E_{\textrm{rel}}  }/\hbar$ is the relative momentum of the $n\text{-}A$ system with the reduced mass $m$.  $\Psi_{000}(\boldsymbol{R}_{\textrm{c.m.}})$ is the ground-state harmonic oscillator function describing $0s$-state c.m. oscillations of the ($A+1$)-body system and $\boldsymbol{R}_{\textrm{c.m.}}$ is the c.m. coordinate.

It is convenient to express $\tilde{h}^{\pm}_{l}(k\xi_{A})$ as an infinite series of oscillator functions \cite{MethodsPaper2025}:
\begin{equation}
\tilde{h}^{\pm}_{l}(k \xi_{A})=\sum_{N_{\textrm{rel}}=l, l+2,...}^{\infty}C^{\pm}_{\frac{N_{\textrm{rel}}-l}{2},l}(k)R_{\frac{N_{\textrm{rel}}-l}{2},l}(\xi_{A}).
\label{eq9}
\end{equation}
Here, the coefficients $C^{\pm}_{(N_{\textrm{rel}}-l)/2,l}(k)$ are given in Refs. \cite{Zaitsev1998, BangHORSE, MethodsPaper2025}.  The quantity $(N_{\textrm{rel}}-l)/2$ is the radial oscillator quantum number of the relative $n\text{-}A$ motion.  The radial oscillator functions $R_{(N_{\textrm{rel}}-l)/2,l}(\xi_{A})$ are parameterized by the oscillator length parameter $b=\sqrt{\hbar / m \Omega}$ where $\Omega$ is the oscillator frequency.  For the Daejeon16 $NN$ interaction, values of $\hbar \Omega=22.5{\text{-}}27.5$ MeV were found to provide good convergence for $n \alpha$ scattering calculations.  Expanding $\tilde{h}^{\pm}_{l}(k\xi_{A})$ in oscillator functions allows the channel function in Eq. \eqref{eq2} to be expressed in terms of the channel-function components $\hat{\mathcal{A}}[[\chi^{\frac{1}{2}}\Psi^{I_{2}}(\boldsymbol{\xi})]_{S}R_{\frac{N_{\textrm{rel}}-l}{2},l}(\xi_{A})Y_{l}(\boldsymbol{\hat{\xi}}_{A})]_{JM_{J}}\Psi_{000}(\boldsymbol{R}_{\textrm{c.m.}})$.

The scattering wave function $\Psi^{\Upsilon}$ satisfies the Schrödinger equation $(H-E)\Psi^{\Upsilon}=0$, where $H$ is the Hamiltonian of the $(A+1)$-body system.  To evaluate the $v+1$ unknowns $\mathbb{S}(E_{\textrm{rel}})$ and \mbox{$b_{q}(E), \hspace{1mm} q=1,...\:,v$}, we solve the set of linear equations \cite{MethodsPaper2025}
\begin{equation}
\langle \bar{\beta_{q}}|H-E|\Psi^{\Upsilon}\rangle=0, \quad q=1,...,v+1,
\label{eqEfrosMethod}
\end{equation}
where $\bar{\beta_{q}}, \hspace{1mm} q=1,...,v+1$ are $v+1$ linearly-independent ($A+1$)-body SRFs.  Equation \eqref{eqEfrosMethod} is equivalent to solving the set of linear equations 
\begin{equation}
\mathbb{A}c=\mathbb{B}. 
\label{EfrosMatrixEquations}
\end{equation}
 Here, the elements of matrix $\mathbb{A}$ are \cite{MethodsPaper2025}
\begin{subequations}
\begin{align}
\mathbb{A}_{q q'}=\langle  {\bar{\beta}_{q}}|H-E|\beta_{q'} \rangle, \quad &\begin{cases}
							q=1,...\, ,v+1, \\ 
							q'=1,...\, ,v, \end{cases}
\label{eq24a}
\\
\mathbb{A}_{q,v+1}= {\langle {\bar{\beta}_{q}}|H-E| \eta^{+} \rangle } ,  \quad &
							{q=1,...\, ,v+1} ,			 
\label{eq24b}
\intertext{the components of vector $\mathbb{B}$ are}
\mathbb{B}_{q}=-\langle  {\bar{\beta}_{q}}|H-E| \eta^{-}  \rangle  , \quad 
							&{q=1,...\, ,v+1}, 
\label{eq24c}
\intertext{and the components of vector $c$ are}
c_q=  \begin{cases}
\phantom{-}b_{q}(E), \quad q=1,...\, ,v, \\
-\mathbb{S}(E_{\textrm{rel}}),  \quad q=v+1. 
\end{cases}
\label{eq24d}
\end{align}
\label{eqMatrixElements}
\end{subequations}
The resonance energies and widths for any partial wave of neutron-nucleus scattering can be found by locating the $S$-matrix pole, which is equivalent to finding the complex energy $E$ satisfying the equation \cite{MethodsPaper2025}
\begin{equation}
\textrm{det}(\mathbb{A})=0.
\label{detEqn}
\end{equation} 
An efficient technique for solving Eq. \eqref{detEqn} in the complex energy plane is presented in Ref. \cite{AMShirokovSSHORSE2018}.

We write the Hamiltonian as \cite{Lanczos2012}
\begin{equation}
\label{eq16}
H=PHP+Qh^{\textrm{ch}}P+Ph^{\textrm{ch}}Q+Qh^{\textrm{ch}}Q,
\end{equation}
where $PHP=H^{A+1}$ is the finite dimensional NCSM Hamiltonian of the ($A+1$)-body composite system and
\begin{equation}
\label{eq18}
h^{\textrm{ch}}=T^{n\text{-}A}+H^{A},
\end{equation}
where $T^{n\text{-}A}$ is the relative $n\text{-}A$ kinetic energy and $H^{A}$ is the internal Hamiltonian of the target nucleus.

 The $P$-space projector is
\begin{equation}
\label{eq20}
P=\hspace{-5ex} \sum_{N =N_{\textrm{min}}^{\textrm{ex}},N_{\textrm{min}}^{\textrm{ex}}+2,...}^{N_{\textrm{max}}^{A+1}} \sum_{\zeta} |N \zeta \rangle \langle N \zeta|,
\end{equation}
where $N$ is the total number of excitation quanta of the ($A+1$)-body system bounded by the minimum total excitation quanta $N_{\textrm{min}}^{\textrm{ex}}$ and the maximum total number of excitation quanta of the ($A+1$)-body system $N_{\textrm{max}}^{A+1}$.  $\zeta$ labels different ($A+1$)-body Pauli-allowed configurations with the same $N$.  $N_{\textrm{min}}^{\textrm{ex}}=0(1)$ for the natural (unnatural) parity state of the ($A+1$)-body system.  The complementary projector $Q$ contains all states with \mbox{$N> N_{\textrm{max}}^{A+1}$} so that $P+Q=1$ and $PQ=QP=0$.

It is convenient to expand the target nucleus wave function with spin $I_{2}$ and total spin projection $M_{2}$  in terms of oscillator functions
\begin{equation}
\Psi^{I_{2},M_{2}}(\boldsymbol{\xi})=\sum_{N_{\textrm{int}}=0,2,... }^{N_{\textrm{max}}^{A}}d_{N_{\textrm{int}}}\Psi_{N_{\textrm{int}}}^{I_{2},M_{2}}(\boldsymbol{\xi}),
\label{eq28}
\end{equation}
where $N_{\textrm{int}}$ is the number of internal excitation quanta of the $A$-body system bounded by the maximum total number of internal excitation quanta of the $A$-body system  $N_{\textrm{max}}^{A}$, $d_{N_{\textrm{int}}}$ is the probability amplitude, and $\Psi_{N_{\textrm{int}}}^{I_{2},M_{2}}(\boldsymbol{\xi})$ are $A$-body components with number of excitation quanta $N_{\textrm{int}}$.  The total number of oscillator excitation quanta of the ($A+1$)-body system is the sum of the number of internal excitation quanta $N_{\textrm{int}}$ and relative quanta $N_{\textrm{rel}}$.

It is also convenient to represent $T^{n\text{-}A}$ in the basis of relative oscillators,
\begin{equation}
\label{eq31}
T^{n\text{-}A}= \hspace{-6ex} \sum_{N_{\textrm{rel}},N'_{\textrm{rel}}=l,l+2,...}^{\infty} | R_{\frac{N_{\textrm{rel}}'-l}{2},l}Y_{l m_{l}}\rangle T_{N_{\textrm{rel}}',N_{\textrm{rel}}}^{n\text{-}A} \langle R_{\frac{N_{\textrm{rel}}-l}{2},l}Y_{l m_{l}}|,
\end{equation}
where $T_{N_{\textrm{rel}}',N_{\textrm{rel}}}^{n\text{-}A}$ are elements of a tridiagonal matrix given in, e.g., Ref. \cite{BangHORSE}, and $H^{A}$ in the basis of the functions $\Psi_{N_{\textrm{int}}}^{I_{2},M_{2}}(\boldsymbol{\xi})$,
\begin{equation}
\label{eq30}
H^{A}= \hspace{-4ex}  \sum_{N_{\textrm{int}}, N'_{\textrm{int}}=0,2,...}^{N_{\textrm{max}}^{A}} | \Psi_{N'_{\textrm{int}}}^{I_{2},M_{2}} \rangle H_{N_{\textrm{int}}',N_{\textrm{int}}}^{A} \langle \Psi_{N_{\textrm{int}}}^{I_{2},M_{2}}|,
\end{equation}
where $H_{N_{\textrm{int}}',N_{\textrm{int}}}^{A}$ are the target nucleus Hamiltonian matrix elements.

We use SRFs
\begin{gather}
 \bar{\beta}_{q}= 
\ \begin{cases} 
    \beta_{q}^{\textrm{Efn}}, & q=1,...,v,  \\ 
    \bar{\beta}_{v+1}, &  q=v+1,
   \end{cases}
\end{gather}
where we set
\begin{multline}
\label{eqBarBetaNmax2}
\bar{\beta}_{v+1}=\hspace{-0.6ex} \hat{\mathcal{A}}[[\chi^{\frac{1}{2}}\Psi_{0}^{I_{2}}(\boldsymbol{\xi})]_{S}R_{\frac{N_{\textrm{max}}^{A+1}+N_{\textrm{min}}-l}{2}+1,l}(\xi_{A})Y_{l}(\boldsymbol{\hat{\xi}}_{A})]_{JM_{J}}
\\
\times \Psi_{000}(\boldsymbol{R}_{\textrm{c.m.}}).
\end{multline}
Here, $\beta_{q}^{\textrm{Efn}}$ are the eigenfunctions of the $(A+1)$-body NCSM Hamiltonian $PHP$ that are obtainable from shell-model codes such as MFDn (see Refs. \cite{MFDnMaris2010,Euro-Par, Aktulga, NCSM-NTSE2014, Shao}) and $N_{\textrm{min}}$ is the minimal Pauli-allowed total oscillator quanta of the ($A+1$)-body system in the state with the considered spin and parity.  We denote the lowest-lying state of the spin-parity under consideration by $\beta_{1}^{\textrm{Efn}}$, the respective first excited state by $\beta_{2}^{\textrm{Efn}}$, etc.

Using the Schrödinger equation for both the $(A+1)$-body system eigenfunctions and the target nucleus ground state, the oscillator-function expansions of $\eta^{\pm}$, Eq. \eqref{eq9}, and $\Psi^{I_{2},M_{2}}(\boldsymbol{\xi})$, Eq. \eqref{eq28}, and the oscillator-function representations of the operators $P$, $Q$, $T^{n\text{-}A}$, and $H^{A}$, we can write the matrix elements of $\mathbb{A}$ and $\mathbb{B}$ given in Eq. \eqref{eqMatrixElements} in terms of $E_{q}$ [which are the $(A+1)$-body nucleus eigenvalues], $E_{A}$, $T_{N_{\textrm{rel}}',N_{\textrm{rel}}}^{n\text{-}A}$, $H_{N'_{\textrm{int}},N_{\textrm{int}}}^{A}$, $d_{N_{\textrm{int}}}$, and the overlaps $\langle \beta_{q}| \hat{\mathcal{A}}[[\chi^{\frac{1}{2}}\Psi_{N_{\textrm{int}}}^{I_{2}}(\boldsymbol{\xi})]_{S}R_{\frac{N_{\textrm{rel}}-l}{2},l}(\xi_{A})Y_{l}(\boldsymbol{\hat{\xi}}_{A})]_{JM_{J}} \hspace{-0ex}  \Psi_{000}(\boldsymbol{R}_{\textrm{c.m.}} )    \rangle $.  The overlaps can be written in terms of shell-model matrix elements using the techniques presented in Ref. \cite{VDEfros2019} plus simplifications from Ref. \cite{MethodsPaper2025} and can therefore be evaluated using inputs from MFDn.  Once the matrix elements of $\mathbb{A}$ and $\mathbb{B}$ are evaluated, the unknowns, $\mathbb{S}(E_{\textrm{rel}})$, $b_{q}(E), \hspace{1mm} q=1,...,v$ are obtained by solving Eq. \eqref{EfrosMatrixEquations}.

In this work, we consider $n\alpha$ scattering.  Therefore, $A=4$, with the target nucleus being the $\alpha$-particle in its ground state and the composite $(A+1)$-body system being $^5\textrm{He}$.  The reduced mass of the $n\alpha$ system is \mbox{$m=751.14 \hspace{1mm} \textrm{MeV}/c^{2}$} and the considered quantum numbers are $I_{1}=1/2$, $I_{2}=0$, \mbox{$S=1/2$}, and $J^{\pi}=3/2^{-}, 1/2^{-}, 1/2^{+}$.  For all $(A+1)$-body states, we set \mbox{$J=M_{J}$}.

Using the phase shifts $\delta_{l}(E_{\textrm{rel}})$, for the partial waves $J^{\pi}=3/2^{-}, 1/2^{-}, 1/2^{+}$, we calculate the total scattering cross section \cite{Ian}
\begin{equation}
\sigma(E_{\textrm{rel}})=\frac{2\pi}{k^{2}}\sum_{\Upsilon} (2J_{\Upsilon}+1)\sin^{2} \delta_{l_{\Upsilon}}(E_{\textrm{rel}}).
\label{eqCS}
\end{equation}
Here, $J_{\Upsilon}$ and $l_{\Upsilon}$ are, respectively, $J$ and $l$ values associated with the partial wave $\Upsilon$.  

       \begin{figure}[t!]
  \includegraphics[width=1\columnwidth]{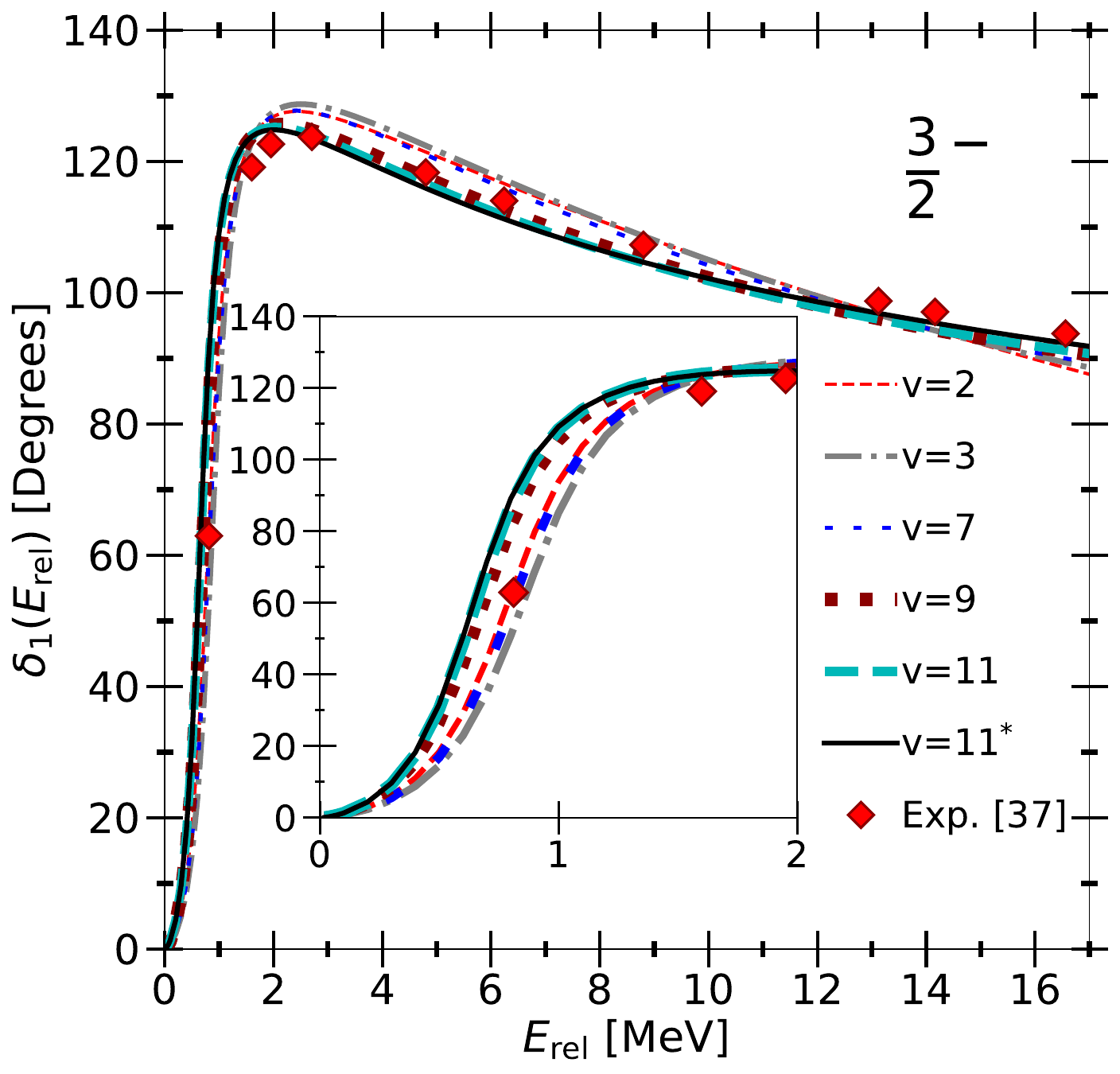}%
\hfill
\caption[]{ Convergence of the $3/2^{-}$ partial wave $n\alpha$ scattering phase shift $\delta_{1}(E_{\textrm{rel}})$ with the number of SRFs $v$ for $^5\textrm{He}$ truncation $N_{\textrm{max}}^{5}=14$, $\alpha$-particle truncation $N_{\textrm{max}}^{4}=6$, and $\hbar \Omega=27.5$ MeV as compared with the results obtained with $v=11$, $N_{\textrm{max}}^{5}=16$, $N_{\textrm{max}}^{4}=6$, and $\hbar \Omega=27.5$ MeV denoted $v=11^{*}$ and with the experimental compilation of data taken from Ref. \cite{Arndt1973Two}. }
 \label{fig1}
\end{figure}

\section{Results}

In Fig. \ref{fig1}, we present the convergence with the number of SRFs $v$ in the $3/2^{-}$ partial wave $n\alpha$ scattering phase shift $\delta_{1}(E_{\textrm{rel}})$.  We use $^5\textrm{He}$ truncation $N_{\textrm{max}}^{5}=14$, $\alpha$-particle truncation $N_{\textrm{max}}^{4}=6$, and \mbox{$\hbar \Omega=27.5$ MeV}.  The SRFs are the $^5\textrm{He}$ eigenfunctions with eigenvalues ranging from \mbox{$E_{1}=1.68$ MeV to $E_{11}=38.2$ MeV} with respect to the $n\alpha$ threshold, with the highest SRF energy lying well beyond the energy range shown in the figure.  For all calculations, $E_{\alpha}=-28.37$ MeV, which is the converged NCSM $\alpha$-particle ground state eigenvalue with the Daejeon16 $NN$ interaction.  To verify that $N_{\textrm{max}}^{5}=14$ is a sufficiently large truncation (the ``$v=11$" case in Fig.~\ref{fig1}), we compare with $\delta_{1}(E_{\textrm{rel}})$ for $v=11$, $N_{\textrm{max}}^{5}=16$, and the same $N_{\textrm{max}}^{4}$ and $\hbar \Omega$ (the ``$v=11^{*}$" case in Fig.~\ref{fig1}) and observe that these results are nearly the same even when expanded in the inset to be distinguishable.

       \begin{figure}[t!]
  \includegraphics[width=1\columnwidth]{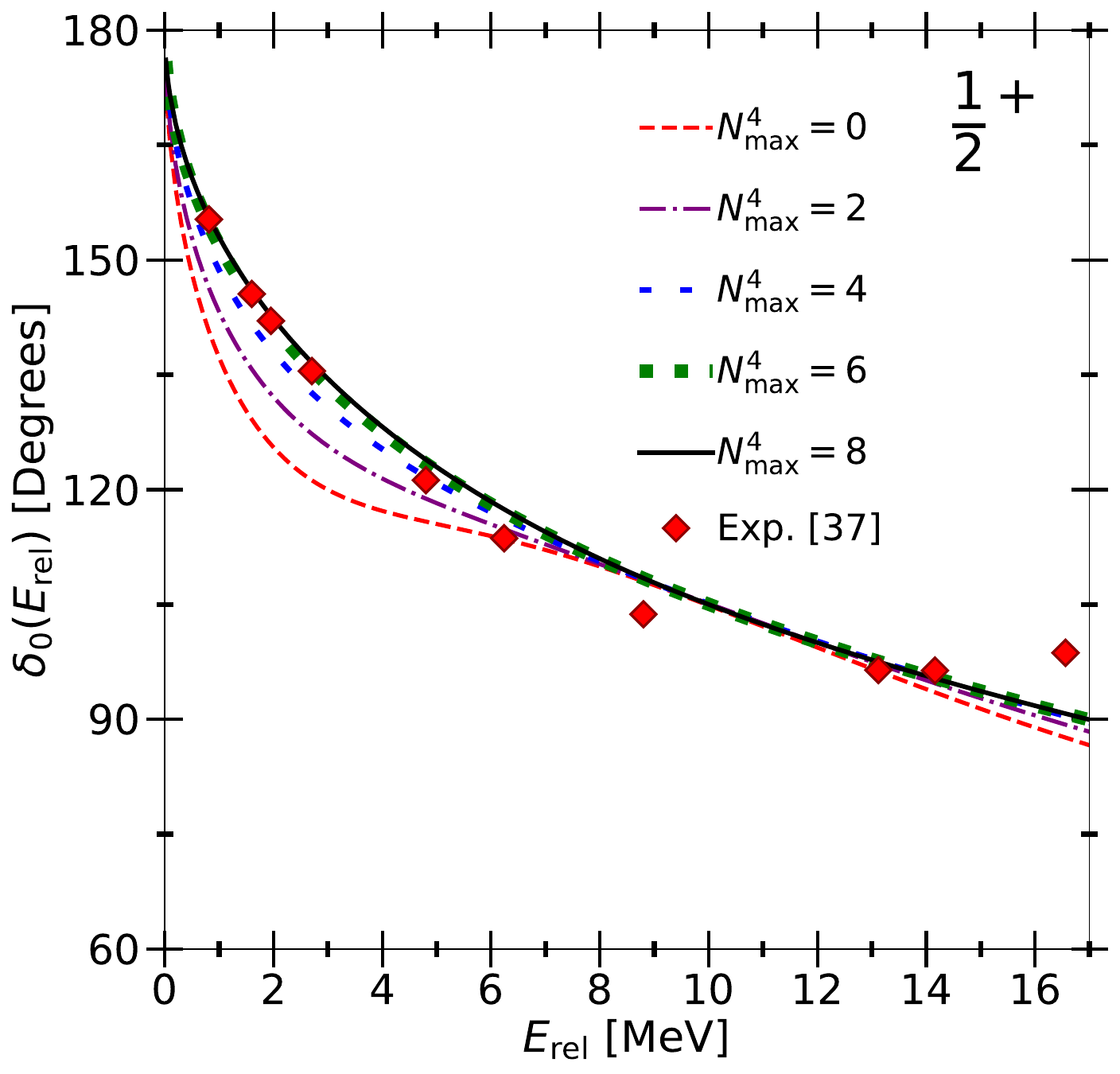}%
\caption[]{  Convergence of the $1/2^{+}$ partial wave $\delta_{0}(E_{\textrm{rel}})$ with $N_{\textrm{max}}^{4}$ for $v=5$, $N_{\textrm{max}}^{5}=13$, and $\hbar \Omega=25$ MeV. }
 \label{fig2}
\end{figure}

We find that $v=11$ SRFs are sufficient for numerical convergence of $\delta_{1}(E_{\textrm{rel}})$ over the energy range considered.  However, good convergence is already achieved with $v=7$ with only a small deviation in the resonance region, while $v=9$ gives convergence over the entire energy range shown.  The results agree with the $N_{\textrm{max}}^{5}=16$  calculation using $v=11$, $N_{\textrm{max}}^{4}=6$, and \mbox{$\hbar \Omega=27.5$ MeV}.  We have also studied the convergence of the $3/2^{-}$ partial wave $\delta_{1}(E_{\textrm{rel}})$ with $N_{\textrm{max}}^{4}>6$ and $v>11$ at $\hbar \Omega=22.5{\text{-}}27.5$ MeV and have found that further extension of $v$ and $N_{\textrm{max}}^{4}$ at $\hbar \Omega=27.5$ MeV does not change the result within the line thickness.  For the limited truncations $v$ and $N_{\textrm{max}}^{5}$ presented, there is only a modest $\hbar \Omega$-dependence that systematically diminishes with increasing truncations.

The resulting $\delta_{1}(E_{\textrm{rel}})$ is in good agreement with the $\delta_{1}(E_{\textrm{rel}})$ obtained from the experimental compilation of data in Ref. \cite{Arndt1973Two} except at energies near resonance, where there is a small deviation.  Using the method of $S$-matrix pole location, we get $E_{r}=0.64$ MeV and \mbox{$\Gamma=0.51$ MeV} for $v=11$, $N_{\textrm{max}}^{5}=14$, $N_{\textrm{max}}^{4}=6$, and \mbox{$\hbar \Omega=27.5$ MeV}.  The experimental values quoted in Ref. \cite{Csoto1997} are \mbox{$E_{r}=0.8$ MeV} and $\Gamma=0.65$ MeV.  The scattering parameters $E_{r}$ and $\Gamma$ are sensitive to the choice of $NN$ interaction.  However, these calculations demonstrate that the Efros method achieves numerical convergence of $\delta_{1}(E_{\textrm{rel}})$ with only a modest number of SRFs $v$ while providing a good description of the $\delta_{1}(E_{\textrm{rel}})$ obtained by fitting the experimental data over most of the energy range considered.

          \begin{figure}[t!]
  \includegraphics[width=1\columnwidth]{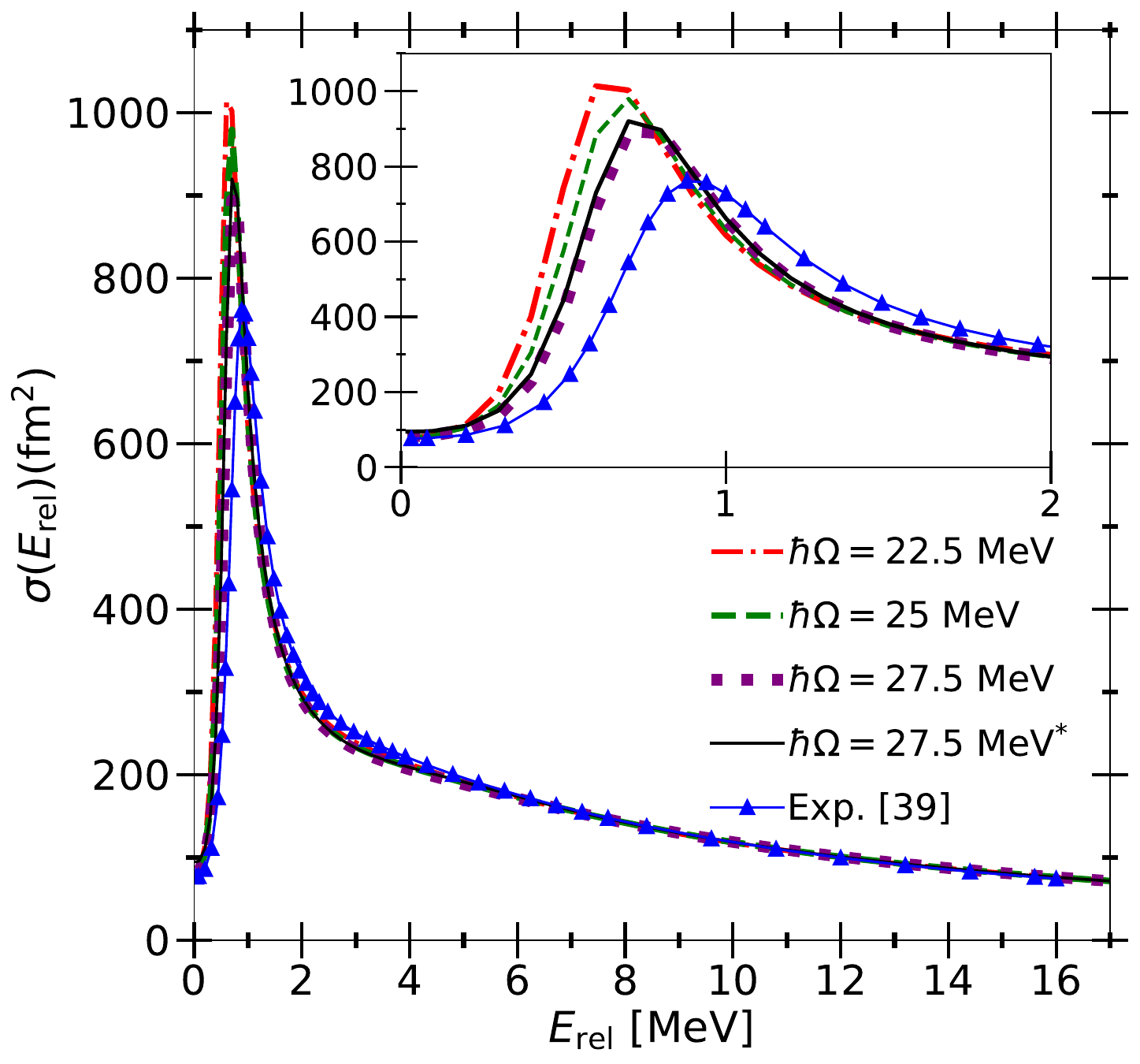}%
\hfill
\caption[]{$n\alpha$ scattering cross section $\sigma(E_{\textrm{rel}})$ for \mbox{$\hbar \Omega=22.5, 25, 27.5$ MeV} for $v=15$, $N_{\textrm{max}}^{5}=13(14)$, and $N_{\textrm{max}}^{4}=6$ as compared with the results obtained with $v=15$, $N_{\textrm{max}}^{5}=15(16)$,  $N_{\textrm{max}}^{4}=6$, and $\hbar \Omega=27.5$ MeV denoted $\hbar \Omega=27.5 \hspace{1mm} \textrm{MeV}^{*}$ and with data based on experimental fits in Ref. \cite{ENDFNNDC}. }
 \label{fig3}
\end{figure}

We have also made similar convergence studies for the $1/2^{-}$ and $1/2^{+}$ partial waves $\delta_{l}(E_{\textrm{rel}})$.  For the $1/2^{-}$ partial wave, the results stabilize with $v=15$ SRFs, $N_{\textrm{max}}^{5}=14$, $N_{\textrm{max}}^{4}=6$, and $\hbar \Omega=20{\text{-}}27.5$ MeV.  For the $1/2^{+}$ partial wave $\delta_{0}(E_{\textrm{rel}})$, convergence is achieved with $v=5$, $N_{\textrm{max}}^{5}=13$, $N_{\textrm{max}}^{4}=6$, and $\hbar \Omega=20{\text{-}}30$ MeV.  As in the $3/2^{-}$ partial wave, the $\delta_{l}(E_{\textrm{rel}})$ results for both the $1/2^{-}$ and $1/2^{+}$ partial waves agree reasonably well with those extracted from experimental data in Ref. \cite{Arndt1973Two}.

In Fig. \ref{fig2}, we present the convergence of the $1/2^{+}$ partial wave $\delta_{0}(E_{\textrm{rel}})$ with $N_{\textrm{max}}^{4}$ for $v=5$, $N_{\textrm{max}}^{5}=13$, and $\hbar \Omega=25$ MeV.  The SRFs are the $^5\textrm{He}$ eigenfunctions with eigenvalues ranging from $E_{1}=6.68$ MeV to \mbox{$E_{5}=27.2$ MeV} with respect to the $n\alpha$ threshold, with the highest eigenvalue lying well beyond the energy range shown in the figure.

Figure \ref{fig2} demonstrates that internal $\alpha$-particle excitations are essential for describing the scattering observables, particularly for the $1/2^{+}$ partial wave.  We see a difference between the converged phase shifts and those obtained with $N_{\textrm{max}}^{4}=0$ at the energies $E_{\textrm{rel}}$ below \mbox{6 MeV}.  This difference diminishes with increasing $N_{\textrm{max}}^{4}$.  We find that allowing for internal $\alpha$-particle excitations up to $N_{\textrm{max}}^{4}=6$ and above gives converged results.  The results also agree reasonably well with the $\delta_{0}(E_{\textrm{rel}})$ extracted from the experimental analysis at higher energies, though the theoretical results tend to deviate from the experimental analysis beyond $E_{\textrm{rel}}=6$ MeV.  It is interesting to note that allowing for the internal $\alpha$-particle excitations modifies $\delta_{0}(E_{\textrm{rel}})$ at low energies while at higher energies the effect of allowing for the $\alpha$-particle excitations is much less pronounced.

 \begin{table*}[t!]
\caption{ $E_{r}$ and $\Gamma$ (in MeV) as compared with other theoretical calculations and experiment.  See the text for details.}
\begin{tabular}{|c|c|c@{\hspace{1pt}\vrule width 1.5pt\hspace{1pt}}@{\hspace{1pt}\vrule width 1.5pt\hspace{1pt}}c|c|}
\hline
 Method & $E_{r}$ ($3/2^{-}$ )  & $\Gamma$ ($3/2^{-}$ ) &    $E_{r}$ ($1/2^{-}$ )  & $\Gamma$ ($1/2^{-}$ ) \\
  \hline
   This paper & 0.59(5)  &  0.49(2)  &     2.87(7) & 6.1(2)  \\
\hline
GFMC \cite{Nollett2007}; AV18 &  1.19   & 1.54       & 1.7 & 4.4  \\
  \hline
  GFMC \cite{Nollett2007}; AV18+UIX  & 1.39    & 1.5      & 2.4 & 5.0  \\
  \hline
 GFMC \cite{Nollett2007}; AV18+IL2  & 0.83  & 0.7  &    2.3 & 5.2  \\
 \hline
NCGSM \cite{Papadimitriou2013}; AV18+Chiral N3LO  & 1.17   &  0.4      & - & -  \\
  \hline
SS-HORSE \cite{AMShirokovSSHORSE2016}; JISP16  & 0.89(7)   &  0.99(7)     & 1.86(9) & 5.46(9)  \\
   \hline
SS-HORSE \cite{AMShirokovSSHORSE2018}; Daejeon16  &  0.68(2)  & 0.52(2)     & 2.45(5)  & 5.07(5)  \\
  \hline
NCSMC \cite{Kravvaris2020}; Chiral N3LO+local $3N$  & 0.756(13)   & 0.663(10)      & 2.32(13)  & 3.84(20)  \\
 \hline
 Exp. \cite{Csoto1997}  &  0.80 &  0.65   &  2.07 & 5.57  \\
  \hline
 \end{tabular}
\label{tab1}
\end{table*}

We have made similar convergence tests for the $1/2^{-}$ partial wave $\delta_{1}(E_{\textrm{rel}})$ and found that the NCSM calculations with at least $N_{\textrm{max}}^{4}=6$ are needed to get converged $\delta_{1}(E_{\textrm{rel}})$ and the main improvements with increasing $N_{\textrm{max}}^{4}$ are also observed at low energies below the resonance.  Increasing $N_{\textrm{max}}^{4}$ to at least 6 quanta also improves the $3/2^{-}$ partial wave $\delta_{1}(E_{\textrm{rel}})$, though the improvements with increasing $N_{\textrm{max}}^{4}$ in this case are much less pronounced than in the $1/2^{+}$ and $1/2^{-}$ partial waves.

In Fig. \ref{fig3}, we present the total $n \alpha$ scattering cross section $\sigma(E_{\textrm{rel}})$  for $\hbar \Omega=22.5, 25, 27.5$ MeV calculated via Eq. \eqref{eqCS}, where we include the $3/2^{-}$, $1/2^{-}$, and $1/2^{+}$ partial waves.  For the $1/2^{+}$ ($3/2^{-}$ and $1/2^{-}$)  partial waves, we take $v=15$, \mbox{$N_{\textrm{max}}^{5}=13(14)$}, and $N_{\textrm{max}}^{4}=6$.  We compare with the results obtained with $v=15$, $N_{\textrm{max}}^{5}=15(16)$, $N_{\textrm{max}}^{4}=6$, and $\hbar \Omega=27.5$ MeV for the $1/2^{+}$ ($3/2^{-}$ and $1/2^{-}$) partial waves and experimental fits in Ref. \cite{ENDFNNDC} based on the experimental results of Refs. \cite{Battat1959,Shamu1964,Rorer1969,Haesner1983}.

 We see some moderate sensitivity of the result to $\hbar \Omega$ around resonance.  However, at energies far from resonance, this sensitivity is minimal.   In addition, we see that the result obtained with $N_{\textrm{max}}^{5}=13(14)$, $N_{\textrm{max}}^{4}=6$, and $\hbar \Omega=27.5$ MeV agrees with the result obtained with $N_{\textrm{max}}^{5}=15(16)$ and the same $N_{\textrm{max}}^{4}$ and $\hbar \Omega$, indicating a good convergence of $\sigma(E_{\textrm{rel}})$.

 Overall, the calculated $\sigma(E_{\textrm{rel}})$ at different $\hbar \Omega$ are in good agreement with experiment over the energy range considered.  Note, however, that the calculations underestimate the resonance location by approximately \mbox{$0.1{\text{-}}0.2$ MeV}.  This discrepancy primarily stems from the inaccuracy in describing experimental fits of the $3/2^{-}$ partial wave $\delta_{1}(E_{\textrm{rel}})$ around resonance as seen in Fig. \ref{fig1}.  Figure \ref{fig3} demonstrates that the converged $\delta_{l}(E_{\textrm{rel}})$ calculations translate into the physical observable $\sigma(E_{\textrm{rel}})$, which is in reasonable agreement with experiment over a broad range of energies.

We present in Table \ref{tab1} resonance energies $E_{r}$ and widths $\Gamma$ in the $3/2^{-}$ and $1/2^{-}$ partial waves of $n\alpha$ scattering by locating within the Efros method respective $S$-matrix poles as discussed in detail in Ref. \cite{MethodsPaper2025}.  We calculate $E_{r}$ and $\Gamma$ at \mbox{$\hbar \Omega=22.5, 25$, and $27.5$ MeV}, taking $v=11$, $N_{\textrm{max}}^{5}=14$, and $N_{\textrm{max}}^{4}=6$.  We take the average of the three results to be the central value. The quoted uncertainties represent the spread associated with the variation of $\hbar \Omega$ over the three values considered.

Our calculated $E_{r}$ and $\Gamma$ agree reasonably well with experiment but display some noticeable discrepancies.  In particular, for the $3/2^{-}$ partial wave, we underestimate $E_{r}$ and $\Gamma$ by about 20\%.  As for the $1/2^{-}$ partial wave, our calculations overestimate $E_{r}$ by approximately 35\% and $\Gamma$ by around 6\%.

We also compare our calculated $E_{r}$ and $\Gamma$ with other theories, namely with Green's function Monte Carlo (GFMC) calculations \cite{Nollett2007} using the AV18 \cite{WiringaAV18}, AV18+UIX \cite{PudlinerUIX}, and AV18+IL2 \cite{PieperIL2} interactions, with NCGSM calculations \cite{Papadimitriou2013} using the AV18+Chiral N3LO \cite{EntemMachleidt2003} interaction, with the SS-HORSE calculations of Ref. \cite{AMShirokovSSHORSE2016} using the JISP16 \cite{JISP16} and of Ref. \cite{AMShirokovSSHORSE2018} using the Daejeon16 $NN$ interactions, and with NCSMC calculations \cite{Kravvaris2020} using the chiral N3LO $NN$ \cite{EntemMachleidt2003}+local 3N \cite{Gazit2009} interactions.

It is not surprising that our results agree most closely with the SS-HORSE calculations of Ref. \cite{AMShirokovSSHORSE2018} with the same Daejeon16 $NN$ interaction.  The differences with the SS-HORSE results are due to the fact that the SS-HORSE calculations in Ref. \cite{AMShirokovSSHORSE2018} include only the lowest $^5\textrm{He}$ eigenvalues corresponding to different $N_{\textrm{max}}^{5}$ and $\hbar \Omega$ whereas the calculations in this work include the excited states.  Moreover, we allow here for internal excitations in the target $^4\textrm{He}$ while the SS-HORSE approach uses the $^4\textrm{He}$ ground state energy obtained with the respective $N_{\textrm{max}}^{4}$ but assumes a simplified $^4\textrm{He}$ wave function where all nucleons fill the lowest oscillator states.  We note also that the calculations of Ref. \cite{AMShirokovSSHORSE2018} required NCSM results obtained for a range of $N_{\textrm{max}}^{5}=14{\text{-}}18$ and $\hbar \Omega=17.5{\text{-}}40$ MeV while in the calculations here we obtained converged results with $N_{\textrm{max}}^{5}=14$ and use a small set of $\hbar \Omega=22.5, 25, 27.5$ MeV needed only to estimate the uncertainty of the results due to the $\hbar \Omega$ variation.  The agreement with the SS-HORSE calculations of Ref. \cite{AMShirokovSSHORSE2018} is better in the $3/2^{-}$ partial wave, where the discrepancy of our calculated $E_{r}$ with that of Ref. \cite{AMShirokovSSHORSE2018} is about 3\% while the resonance widths are consistent within the estimated uncertainties.

Table \ref{tab1} clearly demonstrates that resonance energies and widths are sensitive to the $NN$ and $3N$ interactions.  The best description of the experimental results of Ref. \cite{Csoto1997} for the $3/2^{-}$ and $1/2^{-}$ resonances in $^5\textrm{He}$ was obtained with the AV18 $NN$ + IL2 $3N$ interactions in the GFMC calculations of Ref. \cite{Nollett2007}.

\section{Conclusions}

In this work, we have shown that the version of the Efros method \cite{VDEfros2019} that uses the oscillator functions throughout \cite{MethodsPaper2025} can be successfully combined with the NCSM for realistic many-body scattering calculations.  This work provides the first \textit{ab initio} many-body demonstration of the Efros method.  Using relatively small SRF sets ($v=5\text{-}15$ SRFs) and moderate NCSM bases with $N^{5}_{\textrm{max}}=13$ or 14 in $^5\textrm{He}$ and with $N^{4}_{\textrm{max}}=6$ in $^4\textrm{He}$ calculations, we obtain well-converged phase shifts $\delta_{l}(E_{\textrm{rel}})$ and elastic cross section $\sigma(E_{\textrm{rel}})$ for $\hbar \Omega=22.5{\text{-}}27.5$ MeV, demonstrating that the Efros method is computationally practical for many-body scattering.  We used $S$-matrix pole location to calculate $3/2^{-}$ and $1/2^{-}$ resonance energies $E_{r}$ and widths $\Gamma$ in $^5\textrm{He}$.  The discrepancies in the calculated $E_{r}$ and $\Gamma$ values between theory and experiment and among different theoretical calculations reflect the sensitivity of these quantities to the underlying nuclear interactions.

The extension of this approach to the studies of scattering of charged particles is straightforward \cite{Yanikov2025, YanikovEfros2025}.  The extension to the case of multiple open channels is also possible \cite{VDEfros2019, MethodsPaper2025}.  We plan to use these extensions to enable converged and accurate \textit{ab initio} predictions of various nuclear reactions, in particular, astrophysical reactions as well as reactions involving exotic and unstable nuclei that are extensively investigated experimentally at leading facilities around the world.

\vspace{4mm}

 \section*{ \NoCaseChange{Declaration of generative AI and AI-assisted technologies in the writing process}}
During the preparation of this work, the authors used ChatGPT (OpenAI) in order to improve language, formatting, and readability.  After using these tools/services, the authors reviewed and edited the content as needed and take full responsibility for the content of the publication.

\section*{ \NoCaseChange{Declaration of competing interest}}
The authors declare that they have no known competing financial interests or personal relationships that could have appeared to influence the work reported in this paper.

\section*{\NoCaseChange{Acknowledgements}}
We are thankful to  V. D. Efros and P. Maris for useful discussions.  This work is partially supported by new faculty startup funding by the Institute of Modern Physics, Chinese Academy of Sciences, grant No. E539951SBH, by the Gansu International Collaboration and Talents Recruitment Base of Particle Physics (2023–2027), by the Senior Scientist Program funded by Gansu Province grant No. 25RCKA008, by the Gansu ``Special Program for the Recruitment of Foreign Experts” under grant No. 26RCKA016, by the National Natural Science Foundation of China (No. 12675152), by the Natural Science Foundation of Henan under Grant No. 252300421486, and by the State Key Laboratory of Heavy Ion Science and Technology, Institute of Modern Physics, Chinese Academy of Sciences (Grant No. HST2026CO09).  A portion of the computational resources are provided by the National Energy Research Scientific Computing Center (NERSC), a U.S. Department of Energy Office of Science User Facility located at Lawrence Berkeley National Laboratory, operated under Contract No. DE-AC02-05CH11231 using NERSC awards NP-ERCAP0028672, NP-ERCAP0032689, and NP-ERCAP0037058.  A. M. Shirokov is thankful to the Chinese Academy of Sciences President’s International Fellowship Initiative Program, Grant No. 2026PVA0079, which supported his visits to Lanzhou where a part of this work was done.

 \section*{\NoCaseChange{Data availability}}
 \vspace{-2mm}
\hspace{2mm} Data are available on request to authors.

\end{document}